\documentclass[preprint,aps,superscriptaddress,nofootinbib,tightenlines]{revtex4}
\usepackage{epsfig}
\usepackage{bm}

\def\OMIT#1{{}}

\newcommand{\beq}{\begin{equation}}
\newcommand{\eeq}{\end{equation}}
\newcommand{\bea}{\begin{eqnarray}}
\newcommand{\eea}{\end{eqnarray}}

\newcommand{\benn}{\begin{displaymath}}
\newcommand{\eenn}{\end{displaymath}}

\begin{document}

\begin{figure}[!t]
\vskip -1.5cm
\leftline{
{\epsfxsize=1.8in \epsfbox{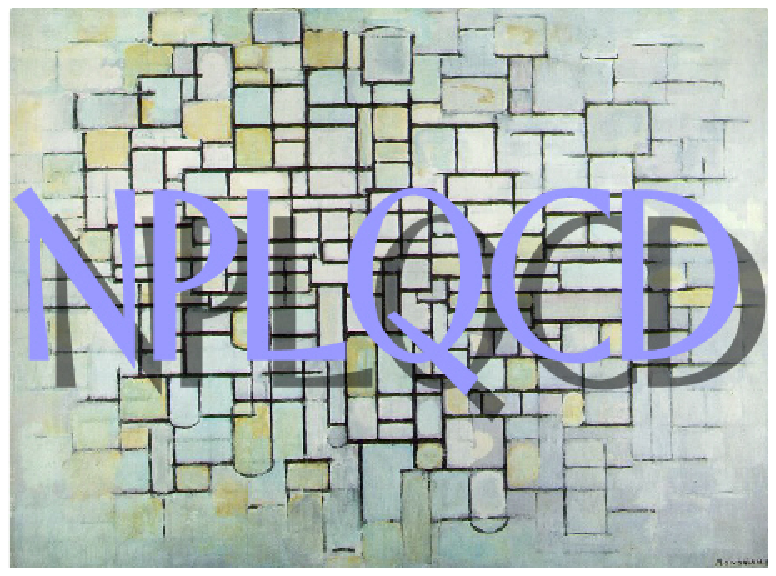}}}
\end{figure}

\preprint{\vbox{
\hbox{UNH-05-04}
\hbox{LBNL-57663}
\hbox{MIT-CTP-3653}
\hbox{CALT-68-2564}
\hbox{NT@UW-05-03}
}}

\vphantom{}
\title{\bf \LARGE  $I=2$ $\pi\pi$ Scattering from Fully-Dynamical Mixed-Action Lattice QCD }
\author{Silas R.~Beane}
\affiliation{Department of Physics, University of New Hampshire,
Durham, NH 03824-3568.}
\affiliation{Jefferson Laboratory, 12000 Jefferson Avenue, 
Newport News, VA 23606.}
\author{Paulo F.~Bedaque}
\affiliation{Lawrence-Berkeley Laboratory, Berkeley,
CA 94720.}
\author{Kostas Orginos}
\affiliation{Center for Theoretical Physics, MIT, Cambridge,
MA 02139.}
\author{Martin J.~Savage}
\affiliation{Center for Theoretical Physics, MIT, Cambridge,
MA 02139.}
\affiliation{California Institute of Technology, Pasadena, CA 91125.}
\affiliation{Department of Physics, University of Washington, 
Seattle, WA 98195-1560.\\
\qquad}
\collaboration{ NPLQCD Collaboration }
\noaffiliation
\vphantom{}
\vskip 0.8cm
\begin{abstract} 
\vskip 0.5cm
\noindent 
We compute the $I=2$ $\pi\pi$ scattering length at pion masses of
$m_\pi\sim 294$, $348$ and $484~{\rm MeV}$ in fully-dynamical lattice
QCD using L\"uscher's finite-volume method.  The calculation is
performed with domain-wall valence-quark propagators on
asqtad-improved MILC configurations with staggered sea quarks
at a single lattice spacing, $b\sim 0.125~{\rm fm}$.
Chiral perturbation theory is used to perform the extrapolation of the
scattering length from lattice quark masses down to the physical
value, and we find $m_\pi a_2 = -0.0426\pm 0.0006 \pm 0.0003\pm
0.0018$, in good agreement with experiment.  The $I=2$ $\pi\pi$
scattering phase shift is calculated to be $\delta = -43\pm 10 \pm
5^o$ at $|{\bf p}|\sim 544~{\rm MeV}$ for $m_\pi\sim 484~{\rm MeV}$.
\end{abstract}
\maketitle

\vfill\eject

\section{Introduction}

\noindent 
The scattering of two pions at low energies is the simplest of all
dynamical strong-interaction processes from the theoretical
standpoint.  As pions are the pseudo-Goldstone bosons associated with
the spontaneous breaking of chiral symmetry, their low-momentum
interactions are constrained by the approximate chiral symmetries of
Quantum Chromodynamics (QCD), and the scattering lengths for
$\pi\pi\rightarrow\pi\pi$ in the s-wave are uniquely predicted at
leading order in chiral perturbation theory ($\chi$-PT).  Higher
orders in the chiral expansion give contributions to these scattering
lengths that are suppressed by powers of $\left( m_\pi/\Lambda_\chi
\right)^2$, where $m_\pi$ is the mass of the pion, and
$\Lambda_\chi\sim 1~{\rm GeV}$ is the scale of chiral symmetry
breaking.  However, part of these higher-order contributions are from
local counterterms with coefficients that are not constrained by
chiral symmetry alone. Lattice QCD is the only known technique with
which one can rigorously calculate these strong-interaction quantities. A
lattice calculation of the $\pi\pi$ scattering length as a function
of the light-quark masses, $m_q$, will allow the determination of 
the relevant counterterms that appear in the chiral lagrangian. Of course, this is
only true for values of $m_\pi$ within the chiral regime where $m_\pi
\ll \Lambda_\chi$. Even though it is likely that in the future lattice
calculations will be performed at the physical values of the
light-quark masses, and while a chiral extrapolation of observables
will not be necessary, the values of the counterterms will be essential
as the chiral lagrangian will still be used to compute
more-complex strong-interaction processes that are too costly to
compute directly on the lattice.  A second important motivation for
studying $\pi\pi$ scattering with lattice QCD is its impact upon
weak-interaction processes (and physics beyond the Standard Model),
such as $K\rightarrow 2\pi$, and the determination of fundamental
constants of nature associated with electroweak physics, 
such as CP violation in the Standard Model.

Lattice QCD calculations are performed on a Euclidean lattice, and the
Maiani-Testa theorem demonstrates that S-matrix elements cannot be
determined from $n$-point Green's functions computed on the lattice at
infinite volume, except at kinematic thresholds~\cite{testa_maiani}.
However, L\"uscher has  shown that by computing the energy levels of
two-particle states in the finite-volume lattice, the $2\rightarrow 2$
scattering amplitude can be recovered~\cite{luscher_formula}.  The
energy levels of the two interacting particles are found to deviate
from those of two non-interacting particles by an amount that depends
on the scattering amplitude and varies inversely with the lattice
spatial volume.  A number of lattice QCD calculations of $\pi\pi$
scattering in the $I=2$ channel have been performed previously using
L\"uscher's method.  For obvious reasons, calculations were initially
performed in quenched
QCD~\cite{Sharpe:1992pp,Gupta:1993rn,Kuramashi:1993ka,Kuramashi:1993yu,Fukugita:1994na,Gattringer:2004wr,Fukugita:1994ve,Fiebig:1999hs,Aoki:1999pt,Liu:2001zp,Liu:2001ss,Aoki:2001hc,Aoki:2002in,Aoki:2002sg,Aoki:2002ny,Juge:2003mr,Ishizuka:2003nb,Aoki:2005uf,Aoki:2004wq}
but in a recent tour-de-force study, the CP-PACS collaboration
exploited the finite-volume strategy to study $I=2$, s-wave $\pi\pi$
scattering in fully-dynamical lattice QCD with two flavors of improved
Wilson fermions~\cite{Yamazaki:2004qb}, with pion masses in the range
$m_\pi\simeq 0.5-1.1~{\rm GeV}$.

A lattice QCD calculation of $\pi\pi$ scattering should fulfill
several conditions: (i) quark masses small enough to lie within
the chiral regime so that the extrapolation to the physical point is
reliable, (ii) a lattice volume large enough to avoid finite-volume effects from 
pions ``going around the world'' but small enough
for the energy shifts to be measurable and (iii) a lattice spacing
small enough so that discretization effects are under control.
Based on these requirements we have performed a fully-dynamical
lattice QCD calculation with two degenerate light quarks and a strange
quark (2+1) with pion masses in the chiral regime, $m_\pi\sim 294~{\rm
MeV}$ (319 configurations), $348~{\rm MeV}$ (649 configurations) and
$484~{\rm MeV}$ (453 configurations).  The configurations we have used
are the publicly-available MILC lattices with dynamical staggered
fermions of spatial dimension $L\sim 2.5~{\rm fm}$, and we have used
domain-wall~\cite{Kaplan:1992bt,Shamir:1993zy,Furman:1994ky,Neuberger:1997fp} 
propagators generated by LHPC~\footnote{ We thank the MILC and LHP collaborations for very
kindly allowing us to use their gauge configurations and light-quark
propagators for this project.  }  at the Thomas Jefferson National
Laboratory (JLab).

This paper is organized as follows.  In Section~\ref{sec:finvol} we
discuss L\"uscher's finite-volume method for extracting hadron-hadron
scattering parameters from energy levels calculated on the lattice.
In Section~\ref{sec:calcdet} we describe the details of our
mixed-action lattice QCD calculation.  We also discuss the relevant
correlation functions and outline our fitting procedures.  In
Section~\ref{sec:matching} we present the results of our lattice
calculation, and the analysis of the lattice data with $\chi$-PT.
In Section~\ref{sec:resdisc} we conclude.

\section{Finite-Volume Calculation of Scattering Amplitudes}
\label{sec:finvol}

\noindent The s-wave scattering amplitude for two particles below
inelastic thresholds can be determined using L\"uscher's
method~\cite{luscher_formula}, which entails a measurement of one or
more energy levels of the two-particle system in a finite volume.  For
two particles of identical mass, $m$, in an s-wave, with zero total
three momentum, and in a finite volume, the difference between the
energy levels and those of two non-interacting particles can be
related to the inverse scattering amplitude via the eigenvalue
equation~\cite{luscher_formula}
\begin{eqnarray}
p\cot\delta(p) \ =\ {1\over \pi L}\ {\bf S}\left(\,\frac{p^2L^2}{4\pi^2}\,\right)\ \ ,
\label{eq:energies}
\end{eqnarray}
where $\delta(p)$ is the elastic-scattering phase shift, and
the regulated three-dimensional sum is
\begin{eqnarray}
{\bf S}\left(\,{\eta}\, \right)\ \equiv \ \sum_{\bf j}^{ |{\bf j}|<\Lambda}
{1\over |{\bf j}|^2-{\eta}}\ -\  {4 \pi \Lambda}
\ \ \  .
\label{eq:Sdefined}
\end{eqnarray}
The sum in eq.~(\ref{eq:Sdefined})
is over all triplets of integers ${\bf j}$ such that $|{\bf j}| < \Lambda$ and the
limit $\Lambda\rightarrow\infty$ is implicit~\cite{Beane:2003da}. 
This definition is equivalent to the analytic continuation of zeta-functions presented 
by L\"uscher~\cite{luscher_formula}.
In eq.~(\ref{eq:energies}), $L$ is the length of the spatial dimension in a 
cubically-symmetric lattice.
The energy eigenvalue $E_n$ and its deviation from twice the rest mass of the particle, $\Delta E_n$,
are related to the center-of-mass momentum $p_n$, a solution of eq.~(\ref{eq:energies}),
by
\begin{eqnarray}
\Delta E_n \ \equiv\ E_n\ -\ 2 m \ =\ 2\;\sqrt{\ p_n^2\ +\ m^2\ }\ -\ 2 m
\ \ \ .
\label{eq:energieshift}
\end{eqnarray}
In the absence of interactions between the particles,
$|p\cot\delta|=\infty$, and the energy levels occur at momenta ${\bf
p} =2\pi{\bf j}/L$, corresponding to single-particle modes in a cubic
cavity.  Expanding eq.~(\ref{eq:energies}) about zero momenta, $p\sim
0$, one obtains the familiar relation~\footnote{ We have chosen to use
the ``particle physics'' definition of the scattering length, as
opposed to the ``nuclear physics'' definition, which is opposite in
sign.}
\begin{eqnarray}
\Delta E_0 &  = &  -\frac{4\pi a}{m  L^3}
\left[\ 1\ +\  c_1 \frac{a}{L}\ +\  c_2 \left( \frac{a}{L} \right)^2 \ \right ]
\ +\ {\cal O}\left({1\over L^6}\right)
\ \ ,
\label{luscher_a}
\end{eqnarray}
with 
\begin{eqnarray}
c_1 & = & {1\over \pi}
\sum_{{\bf j}\ne {\bf 0}}^{ |{\bf j}|<\Lambda}
{1\over |{\bf j}|^2}\ -\   4 \Lambda \
\ =\ -2.8372
\ \ \ ,\ \ \
c_2\ =\ c_1^2 \ -\ {1\over \pi^2} \sum_{{\bf j}\ne {\bf 0}}
{1\over |{\bf j}|^4}
\ =\ 6.37518
\ \ ,
\end{eqnarray}
and $a$ is the scattering length, defined by
\begin{eqnarray}
a & = & \lim_{p\rightarrow 0}\frac{\tan\delta(p)}{p} 
\ \ \ .
\label{eq:scatt}
\end{eqnarray}
For the $I=2$ $\pi\pi$ scattering length, $a_2$, that we consider in
this work, the difference between the exact solution to
eq.~(\ref{eq:energies}) and the approximate solution in
eq.~(\ref{luscher_a}) is much less than $1\%$.  However, in
determining the phase-shift associated with the first excited state on
the lattice, the full eigenvalue equation in eq.~(\ref{eq:energies})
is solved.

\section{Details of the Calculation}
\label{sec:calcdet}

\subsection{Lattices, Actions and Propagators}

\noindent Our computation of the $I=2$ $\pi\pi$ scattering amplitude
consists of a hybrid lattice QCD calculation using staggered sea
quarks and domain-wall valence
quarks~\cite{Schroers:2003mf,Negele:2004iu,Renner:2004ck,Bonnet:2004fr}~\footnote{Technical
aspects of hybridization are discussed in Ref.~\cite{Durr:2003xs} and \cite{Bowler:2004hs}.}.
The parameters of the three sets of $N_f=2+1$
asqtad-improved~\cite{Orginos:1999cr,Orginos:1998ue} MILC
configurations generated with staggered sea
quarks~\cite{Bernard:2001av} that we used in our calculations are
shown in Table~\ref{table:configs}.  In the generation of the MILC
configurations, the strange-quark mass was fixed near its physical
value, $b m_s = 0.050$, (where $b$ is the lattice spacing) determined
by the mass of hadrons containing strange quarks.  The two light
quarks in the three sets of configurations are degenerate, with masses
$b m_l=0.007, 0.010$ and $0.020$.  These lattices were
HYP-blocked~\cite{Hasenfratz:2001hp,Durr:2004as} in order to avoid large residual
chiral symmetry breaking.  We used the domain-wall valence propagators
that had been previously generated by the LHP collaboration on each of
these sets of lattices.  The domain-wall height is $m=1.7$ and extent
of the extra dimension is $L_5=16$.  As this is a mixed-action
calculation, the parameters used to generate the light-quark
propagators have been ``matched'' to those used to generate the MILC
configurations.  This was achieved by requiring that the mass of the pion
computed with the domain-wall propagators be equal (to few-percent precision)
to that of the lightest staggered pion computed from staggered
propagators generated with the same parameters as the given gauge
configuration~\cite{Bernard:2001av}.  The parameters used in the
generation of the domain-wall propagators are shown in
Table~\ref{table:configs}.
\begin{table}[ht]
  \caption{The parameters of the MILC gauge configurations and 
LHPC domain-wall propagators used 
in our calculations. For each propagator the extent of the fifth dimension is $L_5=16$.}
\vskip 0.1in
\begin{tabular}{|c|c|c|c|c|c|c|}
\hline
\label{table:configs}
 Config Set        & \quad Dimensions \quad &\quad $b m_l$ \quad&\quad $b m_s$\quad  
&\quad $b m_{dwf}$\quad  &\quad $10^3 \times b m_{res}$\quad & \# configs  \\
       \hline
MILC\_2064f21b679m007m050 & $20^3\times 64$& 0.007 & 0.05   & 0.0081     & $1.604\pm 0.038$        & 319 \\
MILC\_2064f21b679m010m050 & $20^3\times 64$& 0.010 & 0.05   & 0.0138     & $1.552\pm 0.027$       & 649 \\
MILC\_2064f21b679m020m050 & $20^3\times 64$& 0.020 & 0.05   & 0.0313     & $1.239\pm 0.028$        & 453 \\
  \hline
  \end{tabular}
\end{table}
A Dirichlet boundary condition at the midpoint of the time
direction ($t=32$) has been used in generating the LHPC propagators 
in order to reduce the cost of nucleon matrix-element calculations without
affecting their accuracy\footnote{Note that the source of each propagator was placed 
on the $t=10$ time slice, which led to a maximum of $\Delta t=22$ usable time slices.}. 
Unfortunately, this is not the case for
light-pseudoscalar mesons where the signal is sustained for longer
time intervals, and hence both the systematic errors and the
statistical errors can be improved using correlators of longer time
extent.

\subsection{Correlators, Projections and Fitting Methods}

\noindent 
In order to perform our calculations of the $\pi\pi$ correlation functions
we used the programs {\it Chroma} and {\it QDP++} written at JLab 
under the auspices of SciDAC~\cite{Edwards:2004sx}.
In this programming environment, a few lines of c++ code were required
to construct the two distinct propagator contractions that contribute to $\pi\pi$ interactions
in the $I=2$ channel.
As it is the difference in the energy between two interacting pions and two non-interacting pions that
provides the scattering amplitude, we computed both the one-pion correlation function,
$C_\pi (t)$, and the two-pion correlation function $C_{\pi\pi} (p,t)$, where $t$ denotes the 
number of time-slices between the hadronic-sink and the hadronic-source, and $p$ denotes the 
magnitude of the (equal and opposite) momentum of each pion.

The single-pion correlation function is 
\begin{eqnarray}
C_{\pi^+}(t) & = & \sum_{\bf x}
\langle \pi^-(t,{\bf x})\ \pi^+(0, {\bf 0})
\rangle
\ \ \ ,
\label{pi_correlator} 
\end{eqnarray}
where the summation over ${\bf x}$ corresponds to summing over all the spatial lattice sites, thereby projecting onto 
the momentum ${\bf p}={\bf 0}$ state.
A $\pi^+\pi^+$ correlation function  that projects onto the s-wave state in the continuum limit is
\begin{eqnarray}
C_{\pi^+\pi^+}(p, t) & = & 
\sum_{|{\bf p}|=p}\ 
\sum_{\bf x , y}
e^{i{\bf p}\cdot({\bf x}-{\bf y})} 
\langle \pi^-(t,{\bf x})\ \pi^-(t, {\bf y})\ \pi^+(0, {\bf 0})\ \pi^+(0, {\bf 0})
\rangle
\ \ \ , 
\label{pipi_correlator} 
\end{eqnarray}
where, in eqs.~(\ref{pi_correlator}) and (\ref{pipi_correlator}),
$\pi^+(t,{\bf x}) = \bar u(t, {\bf x}) \gamma_5  d(t, {\bf x})$ is an interpolating field for the  
$\pi^+$.  

Steps were taken to optimize the overlap between the interpolating
fields and the one- and two-pion hadronic states.  First, the
propagators calculated by LHPC, which all have sources centered about
${\bf x}={\bf 0}$, were smeared~\cite{Alexandrou:1992ti,Bonnet:2004fr}
in the neighborhood of ${\bf x}={\bf 0}$ to maximize overlap with the
single-hadron states.  Therefore, in eq.~(\ref{pi_correlator}) and
eq.~(\ref{pipi_correlator}), $\pi^+(0, {\bf 0})$ denotes an
interpolating field constructed from smeared-source light-quark
propagators.  Second, we projected onto two-pion states that are
perturbatively close to the energy eigenstates of interest.
The two-pion states with zero total momentum that transform in
the $A_1$ representation of the cubic group are, in general, linear
combinations of the non-interacting finite-volume eigenstates,
\begin{eqnarray}
| \pi^+ \pi^+ \rangle &  = &   
d_0\ |\pi^+({\bf p}={\bf 0}) \pi^+({\bf p}={\bf 0}) \rangle
\  +\  
d_1\ \sum_{ { {\bf p} L \over 2\pi}= \hat {\bf x}, \hat {\bf y},  \hat {\bf z}}\ 
|\pi^+({\bf p})\pi^+(-{\bf p}) \rangle
\  +\ ... 
\label{eq:pipi_fock} 
\end{eqnarray}
where $d_0$ and $d_1$ are complex coefficients.  In the absence of
interactions, the ground state is given by the first term in
eq.~(\ref{eq:pipi_fock}), the first excited state by the second term,
and so on.  In the regime $|a_2|\ll L$, the effect of the interaction is
small, as is clear from eq.~(\ref{luscher_a}), and each of the terms
in eq.~(\ref{eq:pipi_fock}) are approximate eigenstates.  The momentum
projection in eq.~(\ref{pipi_correlator}) makes it significantly
easier to extract the energies of the interacting eigenstates.

In order to determine the amplitudes and energies of the eigenstates
that our correlation functions contain, we used a variety of fitting
methods.  These include standard covariant fitting, non-covariant
fitting, the effective-mass method, singular-value decomposition,
single-exponential fits, and multiple-exponential fits.  
Table~\ref{table:summary} provides a summary of the results of our calculation.
For each light quark mass we have chosen to extract the necessary quantities by 
fitting the correlation functions over 
the fitting interval given in Table~\ref{table:summary}.
Using a covariance-matrix, $\chi^2$ fitting procedure, along with Jackknifing 
over the lattice configurations, 
we obtain the central values and statistical errors 
(the first error quoted for each quantity) shown 
in Table~\ref{table:summary}.
Varying the fitting interval over reasonable ranges provides an 
estimate of one of the systematic errors associated with
the determination of these quantities. The range of central values 
from a small number of different fitting intervals 
is used to determine the second error associated with 
$m_\pi$, $f_\pi$, $m_\pi/f_\pi$, $\Delta E_0$, $a_2$ , $m_\pi a_2$ 
and $\Delta E_1$, given in Table~\ref{table:summary}.
In the calculation of the remaining quantities in Table~\ref{table:summary}, 
their associated systematic error was estimated by combining 
statistical and systematic errors of the input quantities
in quadrature.
For $m_\pi$, $f_\pi$ and $|{\bf p}|$ given in units of MeV, there 
is an additional systematic error 
arising from the scale setting, which we estimate from the uncertainty 
in the counterterm, $\overline{l}_4$, 
contributing to $f_\pi$, which is discussed subsequently.
\begin{figure}[!ht]
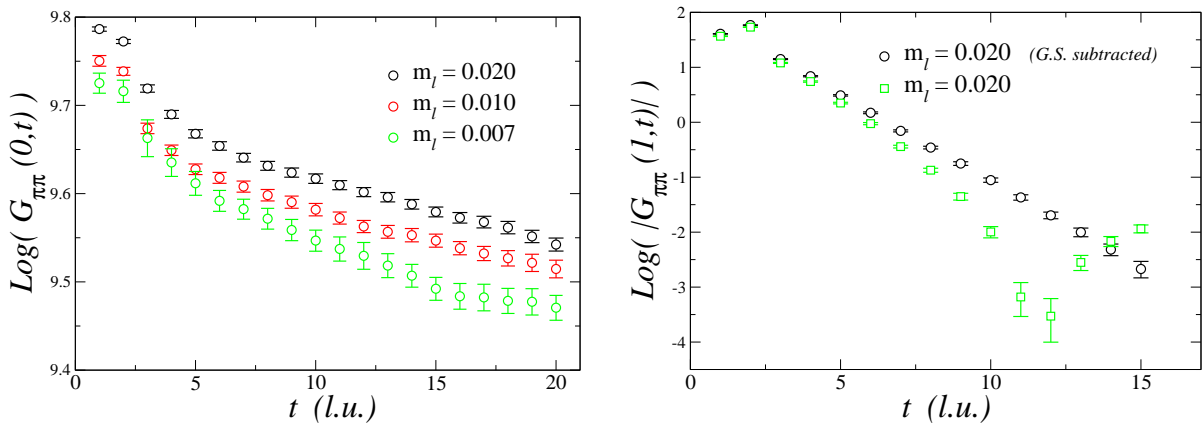

\vskip 0.35in
\centerline{{\epsfxsize=3.0in \epsfbox{fig2.eps}}
\hskip0.2in{\epsfxsize=3.0in \epsfbox{fig3.eps}}}
\vskip 0.15in
\noindent
\caption{Log plots of the ratio of correlation functions, $G_{\pi\pi}$,
defined in eq.~(\ref{ratio_correlator}).
The left panel shows the correlation function $G_{\pi\pi}(0, t)$
for the three sets of MILC configurations, each of which 
is dominated by the $\pi\pi$ ground-state.
The right panel shows the logarithm of the absolute value of $G_{\pi\pi}({2\pi/L}, t)$ 
for the heaviest quark-mass
with and without subtraction of the (small) ground-state component.
}
\label{fig:logcorrgs}
\vskip .2in
\end{figure}

The pion mass was determined by fitting a single exponential to the
pion correlation function, eq.~(\ref{pi_correlator}).  The
point-smeared correlator (point-sink and smeared-source) was found to
be statistically superior to the smeared-smeared correlator (where the
sink is smeared with the same smearing function as the source) both
for the one- and two-pion correlators.  The smeared-smeared correlator
was used only in the determination of the pion decay constant, $f_\pi$
(see below).

In the relatively large lattice volumes that we are using, the energy
difference between the interacting and non-interacting two-pion states
is a small fraction of the total energy, which is dominated by the
mass of the pions.  In order to extract this energy difference we
followed Ref.~\cite{Gupta:1993rn} and formed the ratio of correlation
functions, $G_{\pi\pi}(p, t)$, where
\begin{eqnarray}
G_{\pi\pi}(p, t) & \equiv &
\frac{C_{\pi\pi}(p, t)}{C_{\pi}(t)^2} 
\ \rightarrow \ \sum_{n=0}^\infty\ {\cal A}_n\ e^{-\Delta E_n\ t} 
\  \ ,
\label{ratio_correlator} 
\end{eqnarray}
and the arrow becomes an equality in the limit of an infinite number of
gauge configurations.  In $G_{\pi\pi}(p, t)$, some of the fluctuations
that contribute to both the one- and two-pion correlation functions
cancel, thereby improving the quality of the extraction of the energy
difference beyond what we are able to achieve from an analysis of the individual
correlation functions.

In the $p=0$ case, where we project perturbatively close to the ground
state, the correlator $G_{\pi\pi}(0,t)$, shown in
Fig.~\ref{fig:logcorrgs}, can be fit by a single exponential beyond
the first few time slices, and the ground-state energy difference,
$\Delta E_0$, can be determined quite cleanly.  For the first excited
level, with $p={2\pi/ L}$, the momentum projection in
eq.~(\ref{pipi_correlator}) eliminates all but approximately $10\%$
(in amplitude) of the ground state contribution, which contributes
with opposite sign, as shown in Fig.~\ref{fig:logcorrgs}.  We analyzed
this correlator in two ways. First, by fitting a single exponential to
what remains after subtracting the ground-state contribution
determined at a distant time slice.  Second, by fitting two
exponentials, either keeping the value of the ground state energy
fixed (three parameter fit) or letting that vary also (four parameter
fit).  The difference between these procedures, as well as the
difference coming from the somewhat arbitrary choice of fitting
ranges, is significant only in the value of the two-pion energies and
is incorporated in the estimated systematic error.  The results of our
analysis and the quantities relevant to the determination of the
scattering length are summarized in Table~\ref{table:summary}.

\section{Analysis}
\label{sec:matching}

\noindent A main focus of the analysis of lattice data is the
extrapolation of the lattice values of various physical and unphysical
parameters that hadronic quantities computed on the lattice depend on,
in order to make direct comparison with experimental hadronic
quantities.  Extrapolations in the light-quark masses, the finite
lattice spacing and the lattice volume are currently required.  For
{\it small enough} quark masses and lattice spacings, and large enough
volumes, one can rigorously perform such extrapolations using
low-energy effective field theories.  What defines {\it small enough}
are the dimensionless quantities $b \Lambda_\chi\ll 1$, $m_\pi/
\Lambda_\chi\ll 1$ and $m_\pi L \gg 1$.  The low-energy effective
field theory that can be used to describe the $\pi\pi$ scattering
length is $\chi$-PT supplemented to include the finite lattice
spacing, based on the Symanzik
action~\cite{Sharpe:1998xm,Rupak:2002sm}.

Our calculations have generated pions with masses that are less than
$\sim 500~{\rm MeV}$, which is likely an upper limit to the range of
applicability of the chiral expansion.  We are confident that the
results of the calculations at the lightest two pion masses fall
within the chiral regime and can be analyzed with $\chi$-PT, while the
results of the largest mass point is at the edge of applicability of
$\chi$-PT.

The effect of the finite lattice spacing is not universal and depends
upon the particular discretization used.  In our mixed-action
calculation, both the sea quarks and valence quarks have been included
with actions that have lattice spacing corrections beginning at ${\cal
O}(b^2)$. While $m_{res}\neq 0$ implies ${\cal O}(b)$ corrections,
these effects are exponentially suppressed with the extent of the
fifth dimension and are therefore safely ignored given our current
level of precision. In addition, the sea-quark discretization effects
are further suppressed by the strong coupling constant $\alpha_s$
since the sea-quark action is perturbatively improved.  A
generalization of $\chi$-PT suitable to describe this combination of
mixed-quark-action discretization has recently appeared in
Ref.~\cite{Baer:2005tu}.  Our calculation has been performed at a
single value of the lattice spacing and as such we are unable to
systematically explore the impact of discretization errors on the
determination of the scattering amplitude.  In the analysis that
follows, we have assumed that the lattice-spacing effects are much
smaller than other sources of uncertainty in this calculation, and are
neglected~\footnote{In the $I=0$
$\pi\pi$ channel, unlike the $I=2$ $\pi\pi$ channel, diagrams
involving the double pole in the $\eta'$ propagator, proportional to
$b^2$, lead to modifications to L\"uscher's formula in
eq.~(\ref{luscher_a}) that are enhanced by powers of $L$
\cite{Bernard:1995ez,Golterman:2005xa}.}.

While the power-law finite-volume dependence of the energy of the
two-pion states is used to extract the scattering amplitude using
L\"uscher's method, there are also exponentially suppressed,
finite-volume contributions to the observables resulting from pions
``going around the world''.  These corrections are proportional to
$e^{-m_\pi L}$ and are further suppressed by a factor
$m_\pi^2/\Lambda_\chi^2$, since they can arise only from pion loop
diagrams.  As each of the pion masses that we compute in this work
satisfy $m_\pi L\gg 1$, these exponentially-suppressed finite-volume
contributions are neglected in the analysis.

\subsection{Scale Setting}

\noindent In order to determine quantities in physical units, the
lattice scale must be traded for a physical scale.  In many
calculations the mass of the $\rho$-meson is used.  While this seems
fairly convenient, the light-quark mass dependence of the $\rho$-meson
is controversial (for recent work, see Ref.~\cite{Allton:2005fb}), and
moreover, for light enough pions, the $\rho$-meson becomes unstable,
and therefore determining its mass in the chiral regime requires a
series of calculations of the energy levels of two-pion states in a
finite-volume.  A short-distance length scale is also commonly
used~\cite{Sommer:1993ce}, the Sommer scale, $r_0$.  Such a
short-distance scale depends logarithmically upon the light-quark
masses, as stressed in Ref.~\cite{Grinstein:1996gm}, however lattice
calculations indicate that the numerical size of the light-quark mass
dependence is quite small~\cite{Aubin:2004fs,Aubin:2004wf}, i.e. the
coefficients of unknown operators in the quarkonium chiral lagrangian
are numerically small~\footnote{The coefficients are expected to scale
as $R^3$, where $R$ is the characteristic size of the $\overline{Q}Q$
system, from naive dimensional arguments.}.  In this work the pion
decay constant, $f_\pi$, is used to set the physical length scale.
This has the advantage that: (i) $f_\pi$ is a low-energy parameter
with a well-known dependence on the quark masses (as well as the
lattice volume and lattice spacing) determined by $\chi$-PT; (ii) the
decay constant is easily computed on the lattice with domain-wall
valence quarks.  We calculate $f_\pi$ from pion correlation functions
via~\cite{Blum:2000kn}
\begin{eqnarray}
f_\pi\ =\ \frac{{\cal A_{SP}}}{\sqrt{\cal A_{SS}}}\ \frac{2\,\sqrt{2}\,
(\,m_{dwf}\, +\, m_{res}\,)}{\, m_\pi^{3/2}} \ ,
\label{eq:chiralscsetting8}
\end{eqnarray}
where ${\cal A_{SP}}$ is the amplitude of the one-pion correlation
function resulting from a smeared-source and point-sink, and ${\cal
A_{SS}}$ is the amplitude which results from a smeared-source and
smeared-sink. 
The values of $m_{dwf}$ and $m_{res}$ used in this work can be found in 
Table~\ref{table:configs}.

The chiral expansion of $f_\pi$ is~\cite{Gasser:1983yg}
\begin{eqnarray}
f_\pi & = & f\; \left[\ 1 \ +\  { m^2\over 8\pi^2 f^2}\; 
\overline{l}_4\  + \ \mathcal{O}(m^4)\  \right]\ ,
\label{eq:fchexp}
\end{eqnarray}
where $f$ is the chiral-limit value of the decay constant, $m$ is the
pion mass at leading order in the chiral expansion; $\overline{l}_i =
\log{{\Lambda_i^2}/{m^2}}$ with $\Lambda_i$ an intrinsic scale
that is not determined by chiral symmetry. In what follows we will
denote the physical values of the various parameters with a
$phy$-superscript. One can use eq.~(\ref{eq:fchexp}) to find
\begin{eqnarray}
f_\pi\left(\frac{m_\pi}{f_\pi}\right)\ &=& \  f_\pi^{phy}\;\left[\ 1\  +\  
{1\over 4\pi^2}\; \left[ \left({ m_\pi^{phy}\over f_\pi^{phy}}\right)^2\log\frac{m_\pi^{phy}}{f_\pi^{phy}}\
 -\ \left({ m_\pi\over f_\pi}\right)^2\log\frac{m_\pi}{f_\pi} \right]\ \right. \\ \nonumber 
&&\qquad\qquad\qquad\qquad\left. +\ {1\over 8\pi^2}\;\left[ \left(\frac{m_\pi}{f_\pi}\right)^2\ 
-\ \left(\frac{m_\pi^{phy}}{f_\pi^{phy}}\right)^2 \right]\; 
\overline{l}_4^{\;phy} \  + \ \mathcal{O}\left(\left(\frac{m_\pi^2}{16\pi^2 f_\pi^2}\right)^2\right)\   \right]
\label{eq:yupA}
\end{eqnarray}
as well as other variants of this formula that differ only by terms of
higher order in the chiral expansion. By determining $m_\pi/f_\pi$ in
a lattice calculation, and using the experimental values of $m_\pi$
and $f_\pi$ (and $\overline{l}_4$), the pion-decay constant in the
lattice calculation at the value of the pion mass, $f_\pi
(m_\pi/f_\pi)$, is determined at a given order (in this case
next-to-leading order) in the chiral expansion.  Higher-order
corrections to this scale setting can be determined systematically in
the chiral expansion. We use $f_\pi^{phy}=132~{\rm MeV}$ and
$m_\pi^{phy}=138~{\rm MeV}$.  The uncertainty in the scale-independent
parameter $\overline{l}_{4\;{\it exp}}^{\;phy}=4.4 \pm
0.2$~\cite{Gasser:1983yg,Colangelo:2001df} introduces a source of
systematic error of less than $2\%$ in both the pion mass and decay
constant. The lattice spacings obtained using this method are given in
Table~\ref{table:summary} and may be compared with the determination
by MILC~\cite{Aubin:2004fs}, $b=0.1243\pm 0.0015~{\rm fm}$, using
the Sommer scale. The values of the pion masses and decay constants
in physical units computed in this paper are consistent with the same
quantities computed on the same lattices by MILC in
Ref.~\cite{Aubin:2004fs}. For instance, MILC finds $m_\pi=$ $299.99 \pm
0.32 \pm 3.4$, $356.46 \pm 0.27 \pm 4.0$, $494.27 \pm 0.25 \pm 5.6$ on
the $0.007$, $0.010$ and $0.020$ lattices, respectively, where the
first error is statistical and the second arises from scale setting.
One should keep in mind that (i) results differ at $\mathcal{O}(b^2)$ and
$\mathcal{O}(m_\pi^4)$; (ii) the tuning of the domain-wall fermion
mass to reproduce the lightest staggered pion mass is done to order
$1\%$ accuracy; (iii) Our scale setting procedure includes electromagnetic effects
via $f_\pi^{phy}$ while MILC's does not.

Rather than using the low-energy constant $\overline{l}_4^{\;phy}$ as an input to the scale setting procedure,
one can perform a two-parameter fit of the lattice data in Table~\ref{table:summary} to the form
\begin{eqnarray}
f^{\;l.u.}_\pi\left(\frac{m_\pi}{f_\pi}\right)\ &=& \  b_{\it fit}\;f_\pi^{phy} \;\left[\ 1\  +\  
{1\over 4\pi^2}\; \left[ \left({ m_\pi^{phy}\over f_\pi^{phy}}\right)^2\log\frac{m_\pi^{phy}}{f_\pi^{phy}}\
 -\ \left({ m_\pi\over f_\pi}\right)^2\log\frac{m_\pi}{f_\pi} \right]\ \right. \\ \nonumber 
&&\qquad\qquad\qquad\qquad\left. +\ {1\over 8\pi^2}\;\left[ \left(\frac{m_\pi}{f_\pi}\right)^2\ 
-\ \left(\frac{m_\pi^{phy}}{f_\pi^{phy}}\right)^2 \right]\; 
\overline{l}_{4\;{\it fit}}^{\;phy} \   \right]
\label{eq:blahB}
\end{eqnarray}
where $f^{\;l.u.}_\pi$ is the pion decay constant in lattice units.
This procedure gives $b_{\it fit}=0.1274\pm 0.0007 \pm 0.0003~{\rm fm}$ and
$\overline{l}_{4\;{\it fit}}^{\;phy}=4.412 \pm 0.077 \pm 0.068$ where the first error is statistical and the second is an estimate
of the systematic error. Thus this scale setting procedure is remarkably robust and
consistent. One may wonder about the relevance of the chiral logarithm. Repeating the fitting procedure with
$f^{\;l.u.}_\pi\left({m_\pi}/{f_\pi}\right) =  b_{\it fit}\;f_\pi^{phy} \;\left[ 1 + {1/{8\pi^2}}\;\left[ \left({m_\pi}/{f_\pi}\right)^2
- \left({m_\pi^{phy}}/{f_\pi^{phy}}\right)^2 \right]\; {\cal L}_{\it fit} \right]$ yields
$b_{\it fit}=0.1330\pm 0.0001 \pm 0.0001~{\rm fm}$ and ${\cal L}_{\it fit}=1.407 \pm 0.010 \pm 0.009$, which are not 
consistent with MILC scale setting or the experimental value of $\overline{l}_4^{\;phy}$, respectively. It would appear that
the chiral logarithm is resolved by our data at this order in the chiral expansion.

\subsection{The Scattering Length}

\noindent With small quark masses and momenta, $\pi\pi$ scattering can
be reliably computed in $\chi$-PT. The leading-order result
(equivalent to current algebra) was computed in Ref.~\cite{weinberg},
and the one-loop $\pi\pi$ amplitude was computed in
Ref.~\cite{Gasser:1983yg}.  While this amplitude is now known at the
two-loop level~\cite{Bijnens:1995yn,Bijnens:1997vq}, given our current
lattice data, we choose to analyze our lattice results at one-loop
level. The one-loop expression for the $I=2$ $\pi\pi$ scattering
length is
\begin{eqnarray}
m_\pi a_2\ =\ -{m^2_\pi \over 8\pi f_\pi^2}\; \left[\ 
1\ +\ {3 m_\pi^2\over 16\pi^2 f_\pi^2}\; \left(\ 
\log{\frac{m_\pi^2}{\mu^2}}\ + \ l_{\pi\pi}(\mu)\ 
\right) \right]\ ,
\label{eq:ascattGLwithchexp2}
\end{eqnarray}
where $l_{\pi\pi}(\mu)$ is a linear combination of scale-dependent
low-energy constants that appear in the $\mathcal{O}(p^4)$ chiral
lagrangian~\cite{Gasser:1983yg} (see Appendix~\ref{app:twoloop}).  We
define $l_{\pi\pi} \equiv l_{\pi\pi}(\mu=4\pi f_\pi)$, and therefore
we can simply use the ratio $m_\pi/f_\pi$ computed on the lattice to
determine the scattering length using
eq.~(\ref{eq:ascattGLwithchexp2}).  The difference between using the
lattice $f_\pi$ and a fixed $f_\pi$ in the argument of the logarithm
modifies the scattering length only at higher orders in the chiral
expansion.

The lowest-lying energy eigenvalues in the lattice volume, shown in Table~\ref{table:summary}, 
allow us to determine the $I=2$ $\pi\pi$
scattering lengths at the different light-quark masses via
eq.~(\ref{luscher_a}).  Our results for the scattering lengths, and
other parameters are presented in the summary table, Table~\ref{table:summary}.
The location of the first excited state in the lattice
volume allows, in general, for a determination of the phase-shift at
non-zero values of the pion momentum via eq.~(\ref{eq:energies}).  For
the lattice parameters in these calculations we were able to extract
the $I=2$ $\pi\pi$ phase-shift at one (large) momentum at the largest
quark mass, which is shown in Table~\ref{table:summary}.  For the
two lighter quark masses, the first excited state is very near the
four-pion inelastic threshold, and a simple extraction of the $\pi\pi$
phase-shift is not possible.
\begin{table}[ht]
\caption{The summary table. 
The central value for each quantity is determined by fitting the appropriate correlation
function over the indicated ``Fit Range''.
The first uncertainty is statistical.
The second is an estimate of the systematic error in the fitting process (including varying
the fitting range). For quantities with units of MeV, the third uncertainty is theoretical, and is
due to the uncertainty in the low-energy constant, $\overline{l}_4^{\;phy}$, originating from 
scale-setting.
The chi-square test of fit refers to the extraction of $\Delta E_0$.}
\label{table:summary}
\begin{tabular}{@{}|c | c | c | c |}
\hline
\ \ Quantity \ \ & 
$\quad\qquad m_l=0.007\quad\qquad$& $\qquad\qquad m_l=0.010\qquad\qquad$& $\qquad\qquad m_l=0.020\qquad\qquad$  \\
\hline
\ \ Fit Range  \ \ & 
$\quad\qquad 6-13\quad\qquad$ & $\qquad\qquad 5-15 \qquad\qquad$& $\qquad\qquad 7-15\qquad\qquad$  \\
\hline
$m_\pi$ (l.u.) & $0.1900 \pm 0.0021 \pm 0.002$ & 
	$0.2243\pm 0.0010 \pm 0.0005$ & 
	$0.3131 \pm 0.0012 \pm 0.0017$  \\
$f_\pi$ (l.u.) & $0.0937 \pm 0.0012 \pm 0.001$ & 
	$0.0959 \pm 0.0007 \pm 0.0002$ & 
	$0.1021 \pm 0.0007 \pm 0.0012$    \\
$m_\pi/f_\pi$ & $2.030 \pm 0.040 \pm 0.03$ & 
	$2.338 \pm 0.022 \pm 0.005$ & 
	$3.065\pm 0.024 \pm 0.030$  \\
\hline\hline
\ \ $\chi^2/d.o.f.$  \ \ & 
$\quad\qquad 0.19\quad\qquad$ & $\qquad\quad 0.84 \quad\qquad$& $\qquad\quad 1.03\quad\qquad$  \\
\hline
$\Delta E_0$ (l.u.) & $0.0109 \pm 0.0013 \pm 0.0003$  & 
	$0.0080 \pm 0.0005 \pm 0.0003$ & 
	$0.0073 \pm 0.0007 \pm 0.0004$ \\
$a_2$ (l.u.) & $-1.12 \pm 0.12 \pm 0.02$ & 
	$-0.99 \pm 0.06 \pm 0.04$ & 
	$-1.22 \pm 0.09 \pm 0.07$   \\
$m_\pi a_2$ & $-0.212\pm 0.024 \pm 0.004$ & 
	$-0.222 \pm 0.014 \pm 0.009$ & 
	$-0.38 \pm 0.03 \pm 0.02$  \\
${\cal C}$ & $0.29 \pm 0.16 \pm 0.05$ & 
	$0.021 \pm 0.067 \pm 0.041$ & 
	$0.017 \pm 0.082 \pm 0.057$  \\
\hline\hline
$f_\pi$ (MeV) & $144.7 \pm 0.5 \pm 0.4 \pm 1.0$ & 
	$148.8 \pm 0.3 \pm 0.1 \pm 1.5$ & 
	$158.0 \pm 0.3 \pm 0.4 \pm 2.8$   \\
$m_\pi$ (MeV) & $293.7 \pm 5.9 \pm 4.4 \pm 2.0$ & 
	$347.9 \pm 3.3 \pm 0.8 \pm 3.4$ & 
	$484.4 \pm 3.9 \pm 4.9 \pm 8.5$   \\
$b$ ($10^{-2}$fm) & $12.78 \pm 0.19 \pm 0.18 \pm 0.09\ $ & 
	$12.72 \pm 0.11 \pm 0.09 \pm 0.13$ & 
	$12.75 \pm 0.13 \pm 0.21 \pm 0.23$   \\
\hline
$\Delta E_1$ (l.u.) & $0.482 \pm 0.070 \pm 0.049$ & $0.390 \pm 0.030 \pm 0.035$ & $0.308 \pm 0.009 \pm 0.005$    \\
$|{\bf p}|$ (MeV) & $607 \pm 44 \pm 42 \pm 15$ & $551 \pm 29 \pm 31 \pm 12$ & $544 \pm 14 \pm 12 \pm 10$  \\
$\delta(p)$ (degrees)  & -- & -- & $-43 \pm 10 \pm 5$  \\
\hline
\end{tabular}
\end{table}
\begin{figure}[!ht]
\vskip 0.35in
\centerline{{\epsfxsize=5in \epsfbox{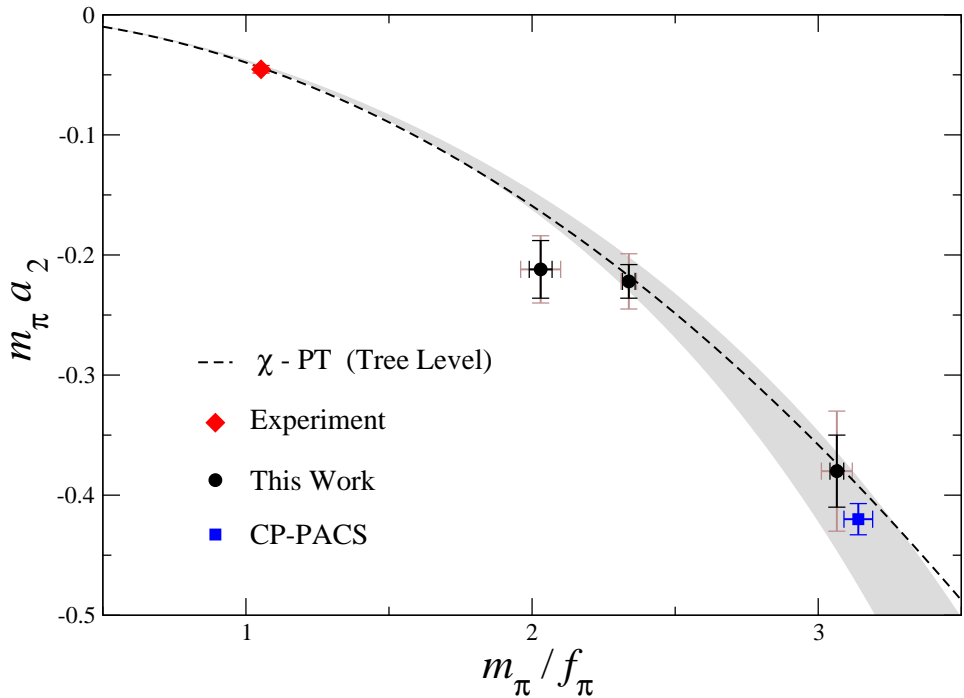}}}
\noindent
\caption{The results of this lattice QCD calculation 
of  $m_\pi a_2$ as a function of $m_\pi/f_\pi$ (ovals) with
statistical (dark bars) and systematic (light bars) uncertainties.
Also shown are the experimental value 
from Ref.~\cite{Pislak:2003sv} (diamond) and the 
lowest quark mass result of the dynamical calculation 
of CP-PACS~\cite{Yamazaki:2004qb} (square). 
The gray band corresponds to a weighted fit to our three data 
points using the one-loop $\chi$-PT formula in
eq.~(\ref{eq:ascattGLwithchexp2}) which gives
$l_{\pi\pi}=3.3 \pm 0.6 \pm 0.3$ 
(the shaded region corresponds only to 
the statistical error). The dashed line is the tree-level
$\chi$-PT result.
} 
\label{fig:extrapolate}
\vskip .2in
\end{figure}
The results of our calculation of the product $m_\pi a_2$ are shown as
a function of $m_\pi/f_\pi$ in Fig.~(\ref{fig:extrapolate}).  In
addition, we have shown the lowest pion mass datum from the dynamical
calculations of the CP-PACS
collaboration~\cite{Yamazaki:2004qb}~\footnote{ We have shown the
CP-PACS data point at the lightest pion mass, and at the smallest
lattice spacing, $\beta=2.10$, and have not attempted to extrapolate
their result to the continuum.  This lattice spacing is comparable to
the one used in this work.}.  The uncertainty in the CP-PACS
measurement is significantly smaller than that of our calculation and
the agreement is very encouraging.  In order to extrapolate $m_\pi
a_2$ to the physical value of $m_\pi/f_\pi$, we performed a weighted
fit of eq.~(\ref{eq:ascattGLwithchexp2}) to the three data points in
Table~\ref{table:summary} and extracted a value of the counterterm
$l_{\pi\pi}$.  As both quantities, $m_\pi a_2$ and $m_\pi/f_\pi$, are
dimensionless there is no systematic uncertainty arising from the
scale setting ($\overline{l}_4$).  We determined that $l_{\pi\pi}=3.3
\pm 0.6 \pm 0.3$, where the first error is statistical and the second
is an estimate of the systematic error (see Fig.~(\ref{fig:extrapolate})). 
This fit of $l_{\pi\pi}$
allows, through eq.~(\ref{eq:ascattGLwithchexp2}), a prediction of the
scattering length at the physical value of the light-quark masses,
which we find to be
\begin{eqnarray}
m_\pi a_2 & = & -0.0426\pm 0.0006 \pm 0.0003\pm 0.0018
\ \ \ .
\label{eq:chiralma}
\end{eqnarray}
The last uncertainty, $\pm 0.0018$, is the largest and 
is an estimate of the systematic error resulting 
from truncation of the chiral expansion of the scattering length~\footnote{As a
consistency check, one may fit the leading-order $\chi$-PT result multiplied by
an arbitrary coefficient:
$m_\pi a_2 = -{\cal A}\;{m^2_\pi \over 8\pi f_\pi^2}$. This yields
${\cal A}=1.058 \pm 0.059 \pm 0.032$
where the first error is statistical and the second is an estimate
of the systematic error. The scattering length at the physical value of the
light-quark masses is, in this case,
$m_\pi a_2  =  -0.0462\pm 0.0026 \pm 0.0014\pm 0.0018$.}.
The two-loop expression for the scattering length~\cite{Bijnens:1997vq,Colangelo:2001df} 
is given by
\begin{eqnarray}
m_\pi a_2\ &=&\ -{m^2_\pi \over 8\pi f_\pi^2}\; \left\{ \ 
1\ +\ {3 m_\pi^2\over 16\pi^2 f_\pi^2}\; \left[\ \log{\frac{m_\pi^2}{\mu^2}}
\ + \ l_{\pi\pi}(\mu )\ \right] \nonumber \right. \\
& & \left. \qquad\qquad + \ {m_\pi^4\over 64\pi^4 f_\pi^4}\; \left[\ 
\frac{31}{6}\,\left(\log{\frac{m_\pi^2}{\mu^2}}\right)^2 \ +\ l^{(2)}_{\pi\pi}(\mu )\; 
\log{\frac{m_\pi^2}{\mu^2}}\ + \ l^{(3)}_{\pi\pi}(\mu )\ \right]
\right\} ,
\label{eq:ascattGLwithchexp2twoloop}
\end{eqnarray}
where $l^{(2)}_{\pi\pi}$ and $l^{(3)}_{\pi\pi}$ are linear combinations of undetermined
constants that appear in the $\mathcal{O}(p^4)$ and $\mathcal{O}(p^6)$  
chiral lagrangians~\cite{Gasser:1983yg,Bijnens:1997vq} (see Appendix~\ref{app:twoloop}).  
It is not possible to provide a meaningful fit of these three undetermined constants
from the few data points we have calculated. While there are estimates of these
low-energy constants from a variety of sources, using these estimates
in our extrapolation to the physical point would amount to trading one
unknown systematic error for another.  In order to estimate the
systematic error due to extrapolation, we first set
$l^{(2)}_{\pi\pi}=l^{(3)}_{\pi\pi}=0$ and refit $l_{\pi\pi}$ keeping
only the double-log piece at two-loop order. We then set
$l^{(3)}_{\pi\pi}=0$ and do a two-parameter fit to $l_{\pi\pi}$ and
$l^{(2)}_{\pi\pi}$. The difference between the extrapolation of the
one-loop expression and the extrapolation of the two-loop expression
with these simplifications gives the estimate of the systematic error
resulting from truncating the chiral expansion. 

In Fig.~(\ref{fig:extrapolate}) we show the unique prediction of
leading order $\chi$-PT for the scattering length (the dashed line),
which is seen to agree remarkably well with both the lattice data and
the physical value. The gray band in Fig.~(\ref{fig:extrapolate})
shows the statistical $1 \sigma$ region.

\begin{figure}[!ht]
\vskip 0.55in
\centerline{{\epsfxsize=5in \epsfbox{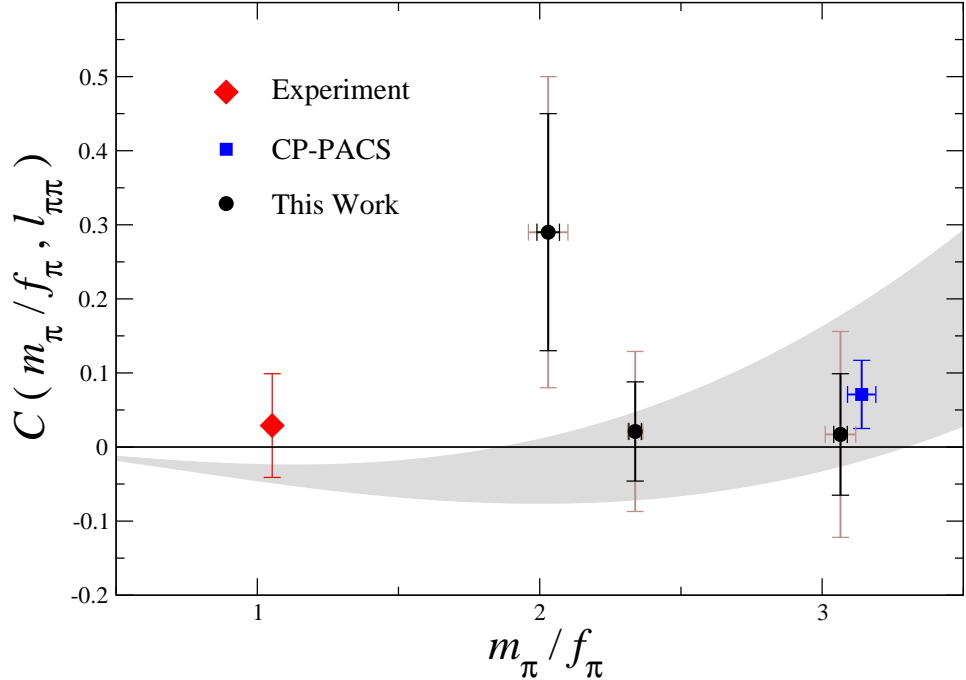}}}
\noindent
\caption{$\mathcal{C}(m_\pi/f_\pi,l_{\pi\pi})$ as a function of $m_\pi/f_\pi$ for
our three lattice data points (ovals) with
statistical (dark bars) and systematic (light bars) uncertainties,
plotted with the experimental value 
from Ref.~\cite{Pislak:2003sv} (diamond) and the lowest-mass dynamical 
CP-PACS~\cite{Yamazaki:2004qb} point (square). 
The gray band corresponds to a fit to our three data points (weighted by
statistical errors) using the one-loop $\chi$-PT formula in
eq.~(\ref{eq:curvefun}).} 
\label{fig:chiralshit}
\end{figure}

In order to better isolate the contributions from higher orders in $\chi$-PT
it is convenient to define a scale-independent ``curvature'' function ${\cal C}$:
\begin{eqnarray}
{\cal C}\left({m_\pi\over f_\pi},l_{\pi\pi}\right)\ \equiv \ 
-{8\pi f_\pi^2\; a_2\over m_\pi}\ - \ 1\ = 
\ {3 m_\pi^2\over 16\pi^2 f_\pi^2}\; \left(\ \log{\frac{m_\pi^2}{16\pi^2 f_\pi^2}}
\ + \ l_{\pi\pi}\ \right) 
\ \ \ ,
\label{eq:curvefun}
\end{eqnarray}
where, by construction, $\mathcal{C}=0$ at tree level.
The values for $\mathcal{C}(m_\pi/f_\pi,l_{\pi\pi})$ 
that we have calculated are 
listed in Table~\ref{table:summary}, and are shown in Fig.~(\ref{fig:chiralshit}) 
as a function of $m_\pi/f_\pi$ together with the experimental point and the CP-PACS result.
A weighted fit of eq.~(\ref{eq:curvefun}) to the results of the lattice calculation
gives: $l_{\pi\pi}=3.3 \pm 0.6\pm 0.2$, consistent with the fit to $m_\pi a_2$.
The gray band in Fig.~(\ref{fig:chiralshit}) shows the $1 \sigma$ region for 
$\mathcal{C}$. The extrapolated value of the scattering length
is obviously the same as in the direct fit to $m_\pi a_2$.

The value of $l_{\pi\pi}$ favored by the fits are such that there is
an almost perfect cancellation 
between the counterterm and the logarithm in eq.~(\ref{eq:ascattGLwithchexp2})
in the range of $m_\pi$ considered.  This
cancellation may be an unfortunate coincidence in this channel. A more
refined lattice QCD calculation is required in order to detect the
predicted chiral curvature. 

The best experimental determination of the $I=2$ scattering length is
obtained through an analysis of $K(e4)$ decays~\cite{Pislak:2003sv},
which gives $m_\pi a_2 = -0.0454 \pm 0.0031 \pm 0.0010 \pm 0.0008$
where the first error is statistical, the second is systematic and the
third is theoretical. This data point is plotted in
Fig.~(\ref{fig:extrapolate}) and in Fig.~(\ref{fig:chiralshit}).  The
two-loop $\chi$-PT ``prediction''~\cite{Colangelo:2001df} is $m_\pi
a_2 = -0.0444 \pm 0.0010$.  Our result is in very good agreement with these
determinations.

\subsection{The Phase Shift}

\noindent
We are only able to extract the phase-shift at one non-zero value of
the pion momentum, and only at the heaviest pion mass, despite having
a relatively clean signal for the first excited state in the lattice
volume for all three sets of MILC configurations.  The reason for this
is that the pion masses are sufficiently light that the first excited
state is at an energy very near the four-pion inelastic threshold on
the two sets of configurations with the lightest pion masses. At
$m_\pi \sim 484~{\rm MeV}$, and for pion momentum $|{\bf p}| \sim
544~{\rm MeV}$, the phase-shift is found to be $\delta=-43\pm 10\pm 5$
degrees.

\section{Conclusions}
\label{sec:resdisc}

\noindent In this paper we have presented the results of a lattice QCD
calculation of the $I=2$ $\pi\pi$ scattering length performed with
domain-wall valence quarks on asqtad-improved MILC configurations with
2+1 dynamical staggered quarks.  The calculations were performed at a
single lattice spacing of $b\sim 0.127~{\rm fm}$ and at a single
lattice size of $L\sim 2.5~{\rm fm}$ with three values of the light
quark masses, corresponding to pion masses of $m_\pi\sim 294, 348$,
and $484~{\rm MeV}$.  We have also presented the phase-shift at the
heavier quark mass.

We have used one-loop $\chi$-PT to fit the combination of
counterterms contributing to $\pi\pi$ scattering at next-to-leading
order to the lattice data and extrapolated in the light-quark masses
down to the physical point.  At the one-loop level we are able to make
a prediction for the value of the scattering length,
$m_\pi a_2=-0.0426\pm 0.0006 \pm 0.0003\pm 0.0018$, which agrees within errors with the experimental value.

\acknowledgments

\noindent 
We thank R.~Edwards for help with the QDP++/Chroma programming
environment~\cite{Edwards:2004sx} with which the calculations
discussed here were performed. We are also indebted to the MILC and
the LHP collaborations for use of their configurations and
propagators, respectively. MJS would like to thank the Center for
Theoretical Physics at MIT and the High-Energy and the Nuclear-Theory
groups at Caltech for kind hospitality during the completion of this
work.  The work of MJS is supported in part by the U.S.~Dept.~of
Energy under Grant No.~DE-FG03-97ER4014.  The work of KO is supported
in part by the U.S.~Dept.~of Energy under Grant No.~DF-FC02-94ER40818.
PFB  was supported in part by the Director, Office of Energy Research,
Office of High Energy and Nuclear Physics, by the Office of Basic
Energy Sciences, Division of Nuclear Sciences, of the U.S.~Department
of Energy under Contract No.~DE-AC03-76SF00098.  The work of SRB is
supported in part by the National Science Foundation under grant
No.~PHY-0400231 and by DOE contract DE-AC05-84ER40150, under which the
Southeastern Universities Research Association (SURA) operates the
Thomas Jefferson National Accelerator Facility.

\vfill\eject

\appendix

\section{The $I=2 $ Scattering Length at Two Loops}
\label{app:twoloop}

\noindent To accuracy $\mathcal{O}(m_\pi^6)$ in $\chi$-PT (two loops), the
scattering length is given by~\cite{Bijnens:1997vq}
\begin{eqnarray}
m_\pi a_2 &=& -\frac{m_\pi^2}{8\pi f_\pi^2}\left\{1-  \frac{m_\pi^2}{8\pi^2 f_\pi^2}\left[ 2+16\pi^2\,{\cal B}\, \right] 
+ \frac{m_\pi^4}{64\pi^4 f_\pi^4}\left[\frac{262}{9}-\frac{22\pi^2}{9}+64\pi^2\,{\cal B}\,\right]\right\}
\label{eq:twoloopB}
\end{eqnarray}
where
\begin{eqnarray}
{\cal B} &=& - \frac{3}{32{ \pi}^{2}}\,\log{\frac{m_\pi^2}{\mu^2}}\ +\ 8\,(\,l_1 +\,l_2\,) + 2\,(\,l_3 -\,l_4\,)\ -\ \frac{3}{32\pi^2}
\nonumber \\
& &\quad\quad +2\,\frac{m_\pi^2}{f_\pi^2}  \left\{ {\vrule height0.86em width0em depth0.86em} \right. \! \!
\frac{1}{16{ \pi}^{2}} \left[ 24\,l_1 + 16\,l_2 + 15 \,l_3 -6\,l_4+\frac{47}{12}\cdot\frac{1}{16\pi^2}\log{\frac{m_\pi^2}{\mu^2}}
+ \frac{1861}{2304{ \pi}^{2}}\right] \nonumber \\
& &\quad\qquad\qquad\quad  +\frac{31}{6}\,\left(\frac{1}{16\pi^2}\log{\frac{m_\pi^2}{\mu^2}}\right)^2 -
\frac{1}{8\pi^2}\left[l_4 + 2(2 l_1+ l_3 +6 l_2) \right]\log{\frac{m_\pi^2}{\mu^2}}  \nonumber \\
& &\quad\quad\qquad\qquad\quad -5 {l_4^2} - 8 {l_3^2} + 4 l_4 \left[8 l_1 + 3 l_3+ 8 l_2 \right] + r_1+ 16 r_4
 \left. {\vrule height0.86em width0em depth0.86em} \right\} \ .
\label{eq:twoloopA}
\end{eqnarray}
The $l_i$ and $r_i$ are scale-dependent low-energy constants that appear in the $\mathcal{O}(p^4)$ and $\mathcal{O}(p^6)$ chiral
lagrangians~\cite{Gasser:1983yg,Bijnens:1997vq}, respectively.

\end{document}